\documentclass[aps,pra,reprint,twocolumn,showpacs,showkeys,floatfix,nobibnotes]{revtex4-2}
\usepackage[utf8]{inputenc}
\usepackage{amsfonts}
\usepackage{amssymb,amsmath}
\usepackage{graphicx,mathptmx,physics,hyperref}
\hypersetup{colorlinks=true, urlcolor=blue, citecolor=blue, linkcolor=blue}
\allowdisplaybreaks 
\begin{document}

\title{Performance of a two-mode coherent superposed channel in continuous-variable quantum teleportation}

\author{Deepak}
\author{Arpita Chatterjee}
\email{Corresponding author: arpita.sps@gmail.com}

\affiliation{Department of Mathematics, J. C. Bose University of Science and Technology, YMCA, Faridabad 121006, Haryana, India}
\begin{abstract}
Glauber's coherent state is denoted by $\ket{\alpha}$ and its two–mode extension is represented by $\ket{\alpha,\beta}$. In this work, we introduce a two–mode superposition operator $A = t\,ab + r\,a^\dagger b^\dagger$ and functioning of it on the two–mode coherent state produces a two–mode coherent superposed quantum state as $\ket{\psi} = (t ab + r a^\dagger b^\dagger)\ket{\alpha,\beta}$. We investigate nonclassicality and quantum non-Gaussianity of this state by means of Wigner distribution and Wigner logarithmic negativity. Once its intrinsic nonclassical and non-Gaussian structure is established, the state is employed as the entangled resource in the Braunstein--Kimble continuous-variable (CV) teleportation protocol. We compute the ideal teleportation fidelity for coherent and squeezed inputs and analyze how the strengths of nonclassicality and non-Gaussianity influence the teleportation efficiency. Our results identify specific parameter regimes where enhanced non-Gaussian features or increased nonclassicality enable fidelities beyond the classical threshold, thereby revealing the operational significance of engineered two–mode quantum states in CV quantum information processing.
\end{abstract}
\maketitle
\section{Introduction}

Nonclassicality and quantum non-Gaussianity are cornerstones of modern quantum optics and continuous-variable (CV) quantum information science. Over the past decade, these concepts have emerged not only as theoretical hallmarks of quantumness but also as practical resources enabling protocols such as entanglement distillation, universal quantum computation, quantum metrology, and secure communication. A state is said to be nonclassical when its Glauber--Sudarshan $P$-function fails to behave as a classical probability distribution \cite{Popov2016}. This implies that no non-negative, well-behaved function exists that expresses the state as a statistical mixture of coherent states. This type of nonclassicality has been widely studied through moment-based witnesses, negativities in the Wigner function, and entanglement potential measures~\cite{PhysRevA.43.492,PuriBook,Deepak2021}. 

Quantum non-Gaussianity captures a distinct and stronger notion of quantumness. In the CV domain, all Gaussian states including squeezed, coherent, and thermal states are fully characterized by first and second-order moments of their quadratures. States lying outside this Gaussian convex hull cannot be simulated by mixtures of Gaussian states, and therefore exhibit quantum features unattainable through Gaussian operations alone. This distinction is operationally crucial because Gaussian tools suffer from significant no-go limitations: they cannot distill Gaussian entanglement~\cite{EisertNogo,FiurasekNogo}, cannot achieve universal quantum computation~\cite{LloydBraunstein}, and cannot protect quantum states from Gaussian noise. As a result, non-Gaussian resources---particularly those generated by photon addition, photon subtraction, and coherent superposition of elementary bosonic operations---have become indispensable while searching for enhanced quantum technologies~\cite{DeepakJPhysB2023, DeepakPhysScr2024}. 
A highly effective strategy for engineering nonclassical and non-Gaussian states involves coherent superposition of photon addition and subtraction. The elementary operation $t a + r a^\dagger$, first proposed in a foundational context by Lee and Nha~\cite{LeeNha}, represents an interference between subtraction and addition processes. This operation is implemented experimentally by erasing which-path information using beam splitters, parametric down-conversion and heralded single-photon detections. Acting on a coherent state, this operator generates a \emph{coherent superposed quantum state} whose nonclassicality can be significantly stronger than that of solely photon-added or photon-subtracted states. Subsequent studies have shown that such engineered states display rich nonclassical features, ranging from higher-order antibunching to strong Wigner negativity and enhanced phase-space correlations~\cite{Deepak2021}.
Based on this development, recent works illustrate that superposition operations and more general nonlinear deformations can dramatically improve the usefulness of non-Gaussian states as entangled resources for quantum teleportation~\cite{DeepakPhysScr2024}. In order to understand the operational relevance of such engineered states, it is helpful to recall the structure of ideal continuous-variable (CV) quantum teleportation. Ideal quantum teleportation refers to the perfect transfer of an unknown quantum state from a sender (Alice) to a receiver (Bob) using pre-shared entangled channel and classical communication, without physically transmitting the state itself ~\cite{Deepak2022,BraunsteinKimble1998}. In the continuous-variable regime, the Braunstein--Kimble (BK) protocol~\cite{BraunsteinKimble1998} describes an ideal teleportation by employing a two-mode entangled resource---typically a two-mode squeezed vacuum---together with a balanced beam-splitter interaction and homodyne detection. Alice mixes the unknown input state with her half of the entangled pair and performs joint quadrature measurements, whose classical outcomes are communicated to Bob. Bob then applies a phase-space displacement conditioned on Alice’s measurement results, thereby reconstructing the input state at his location. In the \emph{ideal} limit of infinite entanglement (equivalently, infinite squeezing), the reconstructed output becomes identical to the input state, yielding unit fidelity. For realistic cases, the fidelity necessarily drops below unity, and only genuine quantum resources can surpass the classical limit of $F=1/2$. Strongly nonclassical and non-Gaussian entangled resources can significantly enhance teleportation performance beyond classical bounds, making ideal CV teleportation a fundamental benchmark for evaluating effectiveness of practical quantum channels and engineered non-Gaussian optical resources.  For instance, photon-added and photon-subtracted displaced Fock states, when mixed with vacuum at a symmetric beam-splitter, form two-mode entangled non-Gaussian channels that outperform Gaussian channels in ideal CV teleportation~\cite{DeepakPhysScr2024}. These states can exhibit enhanced entanglement, stronger Einstein-Podolsky-Rosen (EPR) correlations, and improved teleportation fidelity, supporting the operational importance of engineered non-Gaussian quantum states.

In light of this, the present work introduces a two-mode superposition operator $A = t ab + r a^\dagger b^\dagger$, which serves as a simple two-mode analogue of the single-mode superposition $t a + r a^\dagger$~\cite{Deepak2021b}. This operator annihilates or creates photons in pairs across the two modes simultaneously, enabling correlated nonlinear transformations beyond those achievable with independent single-mode operations. Applying $A$ to the two–mode coherent state $\ket{\alpha,\beta}$ results a family of two-mode coherent superposed quantum states (TMCSQS) of the form $\ket{\psi} = \bigl( t ab + r a^\dagger b^\dagger \bigr) \ket{\alpha,\beta}$ that generalizes former structures in a nontrivial and physically meaningful way. The structure introduced by $ab$ and $a^\dagger b^\dagger$ is particularly appealing because these operators naturally induce correlated excitations across modes, analogous to two-mode squeezing or parametric down-conversion, yet with a coherent superposition that produces additional non-Gaussian interference effects.

In this article, we thoroughly investigate the intrinsic quantum properties of the state $\ket{\psi}$. We analyze its Wigner function~\cite{Royer1977} and quantify its nonclassicality and non-Gaussianity through Wigner logarithmic negativity.
The analysis helps identifying parameter regimes where the state exhibits strong nonclassicality, enhanced non-Gaussianity, or simultaneously strong manifestations of both. We also evaluate the operational power of the generated state as a quantum resource~\cite{Lee2009,Deepak2022}. For this, we employ $\ket{\psi}$ as the entangled channel in the Braunstein--Kimble CV teleportation protocol. Using the characteristic-function formalism, we calculate the ideal teleportation fidelity for coherent and squeezed input states and compare the results with the classical limit ($F=1/2$).
Our analysis reveals clear functional links between nonclassicality, non-Gaussianity, and teleportation performance.

The structure of the paper is as follows. Section~2 presents the theoretical framework for the generated two-mode coherent superposed quantum state. Section~3 investigates the state’s nonclassicality and non-Gaussianity through its phase-space properties and quantitative measures. Section~4 applies TMCSQS as an entangled teleportation channel and evaluates the corresponding fidelity. Section~5 summarizes and discusses the findings of the study.
\section{State of Interest}

In this section, we describe the quantum state that is considered in the subsequent analysis. We begin with the well-known Glauber coherent state $\ket{\alpha}$ \cite{Glauber1963}, defined as the right eigenstate of the bosonic annihilation operator satisfying $a\ket{\alpha} = \alpha \ket{\alpha}$. Its two-mode generalization is given by the tensor product $\ket{\alpha,\beta} = \ket{\alpha}\otimes\ket{\beta}$, which represents a separable classical state in phase space. Since coherent state possesses positive, Gaussian Wigner functions \cite{Royer1977} and mimics the behavior of classical electromagnetic fields, it serves as a natural source to generate nonclassical and non-Gaussian states through suitable nonlinear or interference-based quantum operations \cite{Safaeian2011,Manko1997,RomanAncheyta2011,Roy2000,MatosFilho1996,Recamier2008}.

Motivated by prior research on coherent superposition of photon addition and subtraction in single-mode systems, we introduce a two-mode analogue of such an operation \cite{SantosSanchez2011,Sivakumar1999,RomanAncheyta2013,Drummond1980,Roknizadeh2004,SantosSanchez2012}. The elementary single-mode superposition operator $t a + r a^\dagger$ produces strong nonclassicality by coherently mixing photon annihilation and creation processes. To extend this idea to a parallel two-mode setting, we consider the operator
\begin{equation}
A = t\,ab + r\,a^\dagger b^\dagger,
\end{equation}
where $a$ ($a^\dagger$) and $b$ ($b^\dagger$) denote the annihilation (creation) operators of mode 1 and mode 2, respectively. The coefficients $t$ and $r$ are complex parameters that satisfy a normalization constraint, typically $t^{2} + r^{2} = 1$, ensuring that the relative contribution of the two competing processes is well defined. The operator $ab$ tales off one photon from each mode simultaneously while its counterpart $a^\dagger b^\dagger$ includes one photon in each mode. The coherent superposition of these two operations induces correlations that go beyond those of standard linear Gaussian processes such as beam-splitter mixing or two-mode squeezing. In particular, the term $a^\dagger b^\dagger$ resembles the photon-pair creation process in parametric down-conversion while the $ab$ term introduces correlated pair-annihilation. Their coherent superposition allows for quantum interference between these two paths, producing richer nonclassical and non-Gaussian structures in the resulting quantum state.

Applying the operator $A$ to the initial two-mode coherent state $\ket{\alpha,\beta}$ 
yields the two-mode coherent superposed quantum state as
\begin{equation}
\label{eq2}
\ket{\psi} 
    = N^{1/2}A \ket{\alpha,\beta}
    = N^{1/2}\bigl( t\, ab + r\, a^\dagger b^\dagger \bigr)\ket{\alpha,\beta}.
\end{equation}
with $N$ as the normalization constant given by
 $$N^{-1}=|\alpha\beta|^2+t^2(1+|\alpha|^2+|\beta|^2)+2rt\Re(\alpha^2\beta^2).$$
Since $ab$ and $a^\dagger b^\dagger$ act distinctly on coherent states, the resulting state inherits inputs both from photon-pair subtraction and photon-pair addition. The interplay between these contributions gives rise to complex interference patterns in phase space, potentially leading to features such as Wigner function negativity, enhanced quadrature correlations, higher-order nonclassical effects, and strong deviations from Gaussianity.

For a detailed calculation, it is useful to describe the action of the individual terms on the two-mode coherent state as
\begin{align}
ab \ket{\alpha,\beta} &= \alpha \beta \ket{\alpha,\beta}, \\
a^\dagger b^\dagger \ket{\alpha,\beta} 
    &= \sum_{m,n=0}^{\infty} 
       \frac{\alpha^{m} \beta^{n}}{\sqrt{m!n!}} 
       \sqrt{(m+1)(n+1)} \ket{m+1,n+1}.
\end{align}
Here the first term preserves the coherent-state structure up to a multiplicative factor, while the second term generates a nontrivial two-mode excitation that breaks Gaussianity. The resulting state $\ket{\psi}$ is therefore inherently non-Gaussian and is expected to exhibit nonclassical features even though the input state is classical. 

The two-mode coherent superposed quantum state \eqref{eq2} is the main focus in the rest of this study. In foillowing sections, we analyze its nonclassical properties using the Wigner function, quantify its non-Gaussianity, and evaluate its usefulness as an entangled resource in ideal continuous-variable quantum teleportation.

\section{Wigner Function as a Witness of Nonclassicality and Non-Gaussianity}

Wigner quasiprobability distribution offers a comprehensive phase–space representation of quantum states, enabling precise detection of nonclassical and non-Gaussian features. For a two–mode state characterized by the density operator $\rho$, the Wigner function is defined as \cite{DeepakJPhysB2023,DeepakPhysScr2024,EisertNogo,FiurasekNogo,Deepak2023b,Barbieri2010,Wenger2004}
\begin{widetext}
\begin{equation}
W(\gamma,\gamma^{*};\delta,\delta^{*})=\frac{4}{\pi^{4}}\exp\!\left(2|\gamma|^{2}+2|\delta|^{2}\right)\int d^{2}\lambda\, d^{2}\mu\,\langle -\lambda,-\mu|\rho|\lambda,\mu\rangle\exp\!\left[-2(\gamma^{*}\lambda-\gamma\lambda^{*})-2(\delta^{*}\mu-\delta\mu^{*})\right]
\end{equation}
\end{widetext}
where $\ket{\gamma,\delta}$ is a two-mode coherent state. For Gaussian states, the Wigner function displays a strictly Gaussian profile completely determined by the first and second-order statistical moments. Any deviation immediately signals non-Gaussianity. The Wigner function of the considered state can be expressed as
\begin{widetext}
\begin{align}
\label{wfexp}
W(\gamma,~\delta)&=\frac{4N}{\pi^{4}}\exp\!\left(2|\gamma|^{2}+2|\delta|^{2}-|\alpha|^2-|\beta|^2\right)\int d^{2}\lambda\, d^{2}\mu\,(r\alpha\beta+t\lambda^*\mu^*)(r\alpha^*\beta^*+t\lambda\mu)\nonumber\\&\times\exp\!\left(-|\lambda|^2-|\mu|^2+\alpha^{*}\lambda+\beta^{*}\mu-\alpha\lambda^{*}\allowbreak-\beta\mu^{*}-2(\gamma^{*}\lambda-\gamma\lambda^{*})-2(\delta^{*}\mu-\delta\mu^{*})\right)\nonumber\\&
=4N\exp\!\left(2|\gamma|^{2}+2|\delta|^{2}-|\alpha|^2-|\beta|^2\right)\left(r^2|\alpha\beta|^2+t^2\frac{\partial}{\partial u}\frac{-\partial}{\partial u^*}\frac{\partial}{\partial v}\frac{-\partial}{\partial v^*}+rt\left(\alpha^*\beta^*\frac{-\partial}{\partial u^*}\frac{-\partial}{\partial v^*}+\alpha\beta\frac{\partial}{\partial u}\frac{\partial}{\partial v}\right)\right)\nonumber\\&\times\frac{1}{\pi^{4}}\int d^{2}\lambda\, d^{2}\mu\,\exp\!\left(-|\lambda|^2+a\lambda-a^*\lambda^{*}\right)\exp\!\left(-|\mu|^2+b\mu-b^*\mu^{*}\right)~~~~\text{with $u=\alpha^{*}-2\gamma^{*}$ and $v=\beta^{*}-2\delta^{*}$}\nonumber\\&
=4N\exp\!\left(-2|\alpha-\gamma|^{2}-2|\beta-\delta|^{2}\right)\Big\{r^2|\alpha\beta|^2+t^2(1-|\alpha-2\gamma|^2)(1-|\beta-2\delta|^2)+2rt\Re\Big(\alpha\beta (\alpha-2\gamma)(\beta-2\delta)\Big)\Big\}.
\end{align}
\end{widetext}
The presence of nonclassicality is revealed when the Wigner function fails to behave like a classical probability distribution. In particular, if the Wigner function attains negative values in some region of phase space, the associated state is said to exhibit Wigner negativity, a direct signature of nonclassical behavior. This criterion, often referred to as \emph{Wigner function witness of nonclassicality}, provides a necessary and sufficient condition for nonclassicality of a wide range of single and two-mode states. For many experimentally relevant states including photon-added, photon-subtracted, and coherent superposition engineered states, the degree of Wigner negativity associates strongly with other nonclassical indicators such as antibunching, squeezing, and higher-order amplitude correlations.

For the two-mode coherent superposed quantum state $\ket{\psi}$, the Wigner function comprises interference patterns that arises from the coherent mixing of the photon–pair annihilation and photon–pair creation processes. The term $ab\ket{\alpha,\beta}$ preserves the Gaussian structure of the initial coherent state (up to a multiplicative factor) whereas the operation $a^\dagger b^\dagger$ introduces non-Gaussian excitations. Their coherent superposition produces phase–space fringes and oscillatory behavior that is characteristic of quantum interference. These features frequently drive the Wigner function into negative region, thereby serving as a direct witness of nonclassicality for the engineered state. 
\begin{figure*}
\centering
\includegraphics[scale=1.2]{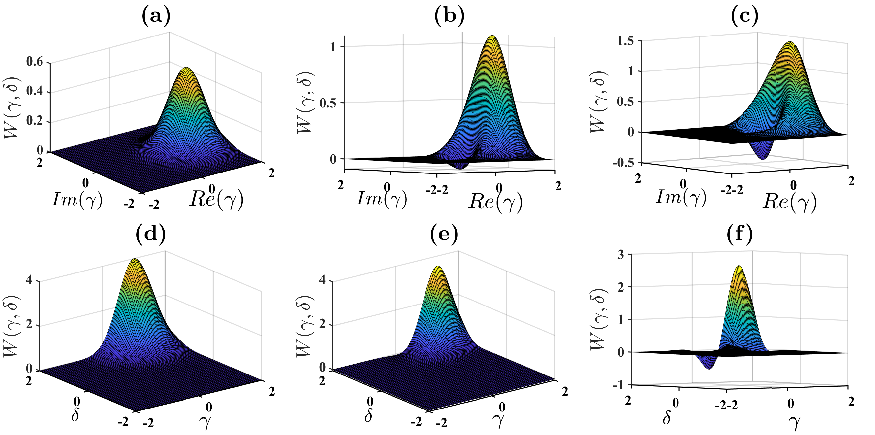}
\caption{A comparison of Wigner function for $\alpha = 1$, $\beta = 2$, illustrating the effects of real and imaginary $\gamma$ in panels (a)--(c) with $\delta = 3$ and $t = 0.01,\,0.56,\,0.98$ respectively, and the combined effects of $\gamma$ and $\delta$ in panels (d)--(f) for the same values of $t$.}
    \label{fig:wigner_compare}
\end{figure*}

The Wigner function of TMCSQS is plotted in Fig.~\ref{fig:wigner_compare} with respect to the real and imaginary parts of the phase-space parameter $\gamma$ in the top row, and with respect to $\gamma$ and $\delta$ in the bottom row.
It is clear that the Wigner function exhibits negative regions and significant deviations from Gaussian behavior, which indicates the presence of nonclassicality and non-Gaussianity in the TMCSQS. Furthermore, both nonclassical and non-Gaussian features become more prominent with increasing values of $t$. Also, enhanced nonclassical and non-Gaussian behavior is observed when $\delta$ is decreasing. The Wigner function attains its extreme values when $\gamma=\delta$. Hence, we conclude that the Wigner function effectively detects the nonclassical and non-Gaussian nature of TMCSQS, which rises as $t$ is increasing but $\delta$ is decreasing.

\begin{figure}
\centering
\includegraphics[scale=0.4]{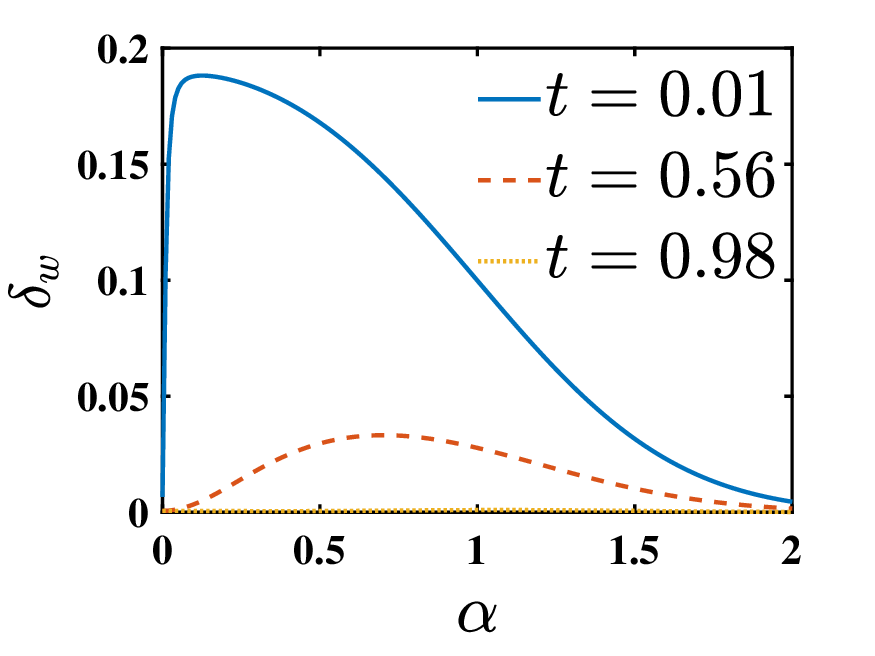}
\caption{Wigner logarithmic negativity with respect to state parameter $\alpha$ with different values of $t$ and $\beta=2$.}
\label{figwln}
\end{figure}

To quantify nonclassicality more systematically, we employ the \emph{Wigner logarithmic negativity}, defined as \cite{PuriBook,EisertNogo,FiurasekNogo,LloydBraunstein}
\begin{equation}
\delta_{W} = \log_2\!\left( \int\!\!\int 
    \left| W(\gamma,\delta) \right| \, d^{2}\gamma\, d^{2}\delta \right),
\label{LogNeg}
\end{equation}
where $\delta_{W} > 0$ indicates the presence of negative regions and larger values of $\delta_{W}$ signify stronger nonclassical nature. Since coherent state possesses purely Gaussian, positive Wigner distribution, any non-zero Wigner negativity in $\ket{\psi}$ arises solely from the coherent action of $ab$ and $a^\dagger b^\dagger$, demonstrating that the superposition operation is responsible for generating significant nonclassicality. The Wigner logarithmic negativity of TMCSQS is plotted in Fig.~\ref{figwln}.
It is clear that $\delta_W$ is positive for all parametric values which indicates that negativity is present in the Wigner function and hence TMCSQS is nonclassical as well as non-Gaussian. Further, Wigner logarithmic negativity shows that nonclassical and non-Gaussian behavior of TMCSQS decreases with respect to both the state parameters $\alpha$ and $t$.

In summary, the Wigner function plays a dual role in characterizing the two-mode coherent superposed quantum state. Its negativity serves as a direct witness of nonclassicality, while deviation from Gaussianity reveals the presence of quantum non-Gaussian features. These properties are crucial for evaluating the usefulness of $\ket{\psi}$ in continuous-variable quantum information protocols, particularly for ideal CV teleportation where strong nonclassicality and non-Gaussianity contribute to enhancing teleportation fidelity.

\section{Ideal Continuous-Variable Teleportation Using TMCSQS Channel}

Continuous-variable (CV) quantum teleportation \cite{Recamier2006,Glauber1963b,SantosSanchez2012,Roknizadeh2004,BraunsteinKimble1998} allows the transfer of an unknown quantum state from a sender (Alice) to a receiver (Bob) using a shared entangled resource and classical communication without physically transmitting the state itself. In the Braunstein--Kimble (BK) protocol \cite{Lvovsky2002,Mandel1979,Puri1996}, the efficiency of teleportation is quantified by the teleportation fidelity \cite{Zavatta2005,Lee1998,Cessa1993,Boiteux1973} which depends critically on the nature of the entangled channel.

In the present work, the entangled resource is chosen to be the two-mode coherent superposed quantum state (TMCSQS), 
where the coherent superposition of photon-pair annihilation and creation processes induces strong nonclassical and non-Gaussian correlations, making TMCSQS a suitable channel for enhanced CV teleportation \cite{Kim2008,Deepak2022,Zavatta2004,Drummond1980,LeeNha}.

Within the BK teleportation protocol, the fidelity for a coherent state input can be expressed using the characteristic-function formalism as
\begin{equation}
F = \frac{1}{\pi} \int d^{2}\lambda \,
\chi_{\mathrm{in}}(\lambda)\chi_{\mathrm{out}}(-\lambda),
\end{equation}
where $\chi_{\mathrm{in}}(\lambda)$ is the characteristic function of the input state and $\chi_{\mathrm{out}}	(\lambda)=\chi_\mathrm{in}(\lambda^{*},\lambda)\chi_\mathrm{res}(\lambda^{*},\lambda)$ denotes the characteristic function of the output state obtained by using the characteristic functions of input and resource channel states.


\subsection{Teleportation of a coherent state input}

For any coherent state input $|\gamma\rangle$, the maximum fidelity attainable using only classical resources is bounded by \cite{LloydBraunstein,Oliveira1990,Raimond2005,Glauber1963a}
\begin{equation}
F_{\mathrm{classical}} = \frac{1}{2}.
\end{equation}
Therefore, achieving $F > 1/2$ is a clear signature of authentic quantum teleportation \cite{RomanAncheyta2013,Sivakumar1999,Mandel1995,SantosSanchez2011,Recamier2008}.

\begin{figure*}
\includegraphics[scale=1.2]{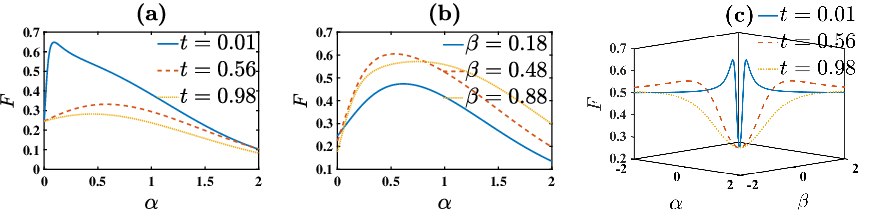}
\caption{Variation of fidelity with coherent state input and with respoect to (a) $\alpha$, $\beta=0.18$, (b) $\alpha$, $t=0.18$, (c) $\alpha$ and $\beta$ and for different values of $t$.}
\label{figics}
\end{figure*}

The fidelity for input coherent state and TMCSQS channel is plotted with respect to $\alpha$ and for different values of $t$ and $\beta$ in Fig.~\ref{figics}.
It is clear that the fidelity is more than the threashold limit of $1/2$. Further, the fidelity decreases with the coherent superposition parameter $t$ and increases with $\alpha$ in the range $\alpha\le 0.5$ and decreases thereafter. Owing to its strong nonclassicality and non-Gaussian correlations, TMCSQS channel yields improved teleportation fidelity to surpass the classical limit for appropriate choices of the superposition parameter $t$ and coherent state parameters $\alpha,~\beta$. In particular, an increased contribution from the photon-pair creation term enhances entanglement and thus teleportation efficiency.

\subsection{Teleportation of a squeezed state input}

Next we consider the teleportation of a single-mode squeezed vacuum state, defined as
\begin{equation}
|\xi\rangle = S(\xi)|0\rangle,
\end{equation}
\begin{equation}
S(\xi) = \exp\!\left[\frac{1}{2}\left(\xi a^{\dagger 2} - \xi^{*} a^{2}\right)\right]
\end{equation}
is the squeezing operator. Squeezed states are inherently nonclassical and exhibit reduced quantum fluctuations in one quadrature, making them highly sensitive probes of teleportation performance.

The teleportation fidelity for a squeezed input state is calculated and plotted in Fig.~\ref{figics}.
\begin{figure*}
\includegraphics[scale=1.2]{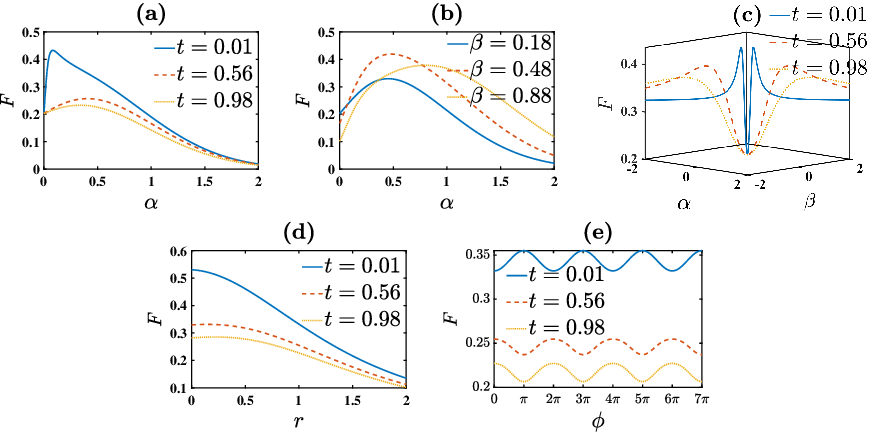}
\caption{Variation of fidelity with squeezed state input ($\xi=re^{i\phi}$) and with respect to (a) $\alpha$ and $\beta=0.18,~r=1,~\phi=0$, (b) $\alpha$ and $t=0.18,~r=1,~\phi=0$, (c) $\alpha$ and $\beta$ and $r=1,~\phi=0$, (d) $r$ and $\alpha=0.48,~\beta=0.18,~\phi=0$, (e) $\phi$ and $\alpha=0.48,~\beta=0.18,~r=1$.}
\label{figiss}
\end{figure*}
It is seen that the fidelity for input squeezed state is just greater than the threshold limit. Further, the fidelity decreases with respect to $\alpha$, $\beta$, $t$, $r$ and oscillatory with respect to $\phi$. 
Teleportation of a squeezed state is more demanding than that of coherent state because Gaussian entangled channels often fail to preserve quadrature squeezing efficiently. In contrast, the non-Gaussian structure of the TMCSQS channel originating from the coherent superposition of photon-pair annihilation and creation processes introduces higher-order correlations that enhance the retention of quadrature information. Thus, TMCSQS channel can achieve improved teleportation fidelity for squeezed input states compared to purely Gaussian resources.

\section{Conclusion}

In this work, we have investigated the nonclassical, non-Gaussian, and quantum communication properties of the two-mode coherent superposed quantum state. By analyzing the Wigner function, Wigner logarithmic negativity, and teleportation fidelity, we have demonstrated that the proposed state exhibits strong nonclassical features and can serve as an efficient quantum resource for continuous-variable quantum information processing.

Our results show that the degree of nonclassicality and phase-space interference can be effectively controlled by means of the superposition parameter $t$, squeezing state amplitude $r$, and coherent amplitudes $\alpha$ and $\beta$. In particular, the superposition parameter $t$ contributes for enhancing of Wigner function negativity that indicates stronger quantum interference effects. The teleportation analysis reveals that TMCSQS can achieve fidelities beyond the classical threshold for a wide range of parameters, confirming its usefulness as a quantum channel. We have also observed that the non-Gaussian nature of the TMCSQS plays a crucial role in improving its performance in quantum communication tasks. Compared with standard Gaussian resources, the proposed state provides enhanced teleportation efficiency in certain parameter regimes, highlighting the advantages of engineered non-Gaussian states. From an experimental viewpoint, TMCSQS may be realized using nonlinear optical processes such as parametric down-conversion combined with conditional measurements and coherent state preparation techniques. This makes the proposed scheme feasible within current photonic quantum technologies.

In conclusion, the two-mode coherent superposed quantum state represents a promising non-Gaussian entangled resource for continuous-variable quantum information protocols. The present study offers a deeper insight into its phase-space structure and operational significance. Future work may focus on investigating its robustness against decoherence and losses, optimizing parametric regimes for practical implementations, and exploring its applications in quantum metrology, quantum cryptography, and quantum networking.

\section{Acknowledgements}
		A. C. acknowledges DST SERB for the support provided through the project number SUR/2022/000899.
\bibliography{ref.bib}
\end{document}